# Post-Radiotherapy PET Image Outcome Prediction by Deep Learning under Biological Model Guidance: A Feasibility Study of Oropharyngeal Cancer Application

Hangjie Ji[1], Kyle Lafata[2,3], Yvonne Mowery[2], David Brizel[2], Andrea L. Bertozzi[1,4], Fang-Fang Yin[2], Chunhao Wang[2]

[1]Department of Mathematics, University of California Los Angeles, Los Angeles, CA 90095
[2]Deparment of Radiation Oncology, Duke University Medical Center, Durham, NC 27710
[3]Deparment of Radiology, Duke University Medical Center, Durham, NC 27710
[4]Mechanical and Aerospace Engineering Department, University of California Los Angeles, Los Angeles, CA 90095

Short Running Title: Biologically-Guided Deep Learning of PET

*Corresponding authors:

Chunhao Wang, Ph.D.
Box 3295, Department of Radiation Oncology
Duke University Medical Center
Durham, NC, 27710, United States
E-mail: chunhao.wang@duke.edu

# Abstract

*Purpose:*

To develop a method of biologically guided deep learning for post-radiation $^{18}$FDG-PET image outcome prediction based on pre-radiation images and radiotherapy dose information.

*Methods:*

Based on the classic reaction-diffusion mechanism, a novel biological model was proposed using a partial differential equation that incorporates spatial radiation dose distribution as a patient-specific treatment information variable. A 7-layer encoder-decoder based convolutional neural network (CNN) was designed and trained to learn the proposed biological model. As such, the model could generate post-radiation $^{18}$FDG-PET image outcome predictions with possible time-series transition from pre-radiotherapy image states to post-radiotherapy states.

The proposed method was developed using 64 oropharyngeal patients with paired $^{18}$FDG-PET studies before and after 20Gy delivery (2Gy/daily fraction) by IMRT. In a two-branch deep learning execution, the proposed CNN learns specific terms in the biological model from paired $^{18}$FDG-PET images and spatial dose distribution as in one branch, and the biological model generates post-20Gy $^{18}$FDG-PET image prediction in the other branch. As in 2D execution, 718/233/230 axial slices from 38/13/13 patients were used for training/validation/independent test. The prediction image results in test cases were compared with the ground-truth results quantitatively.

*Results:*

The proposed method successfully generated post-20Gy $^{18}$FDG-PET image outcome prediction with breakdown illustrations of biological model components. Time-series $^{18}$FDG-PET image predictions were generated to demonstrate the feasibility of disease response rendering. SUV mean values in $^{18}$FDG high-uptake regions of predicted images (2.45±0.25) were similar to ground-truth results (2.51±0.33). In 2D-based Gamma analysis, the median/mean Gamma Index (<1) passing rate of test images was 96.5%/92.8% using 5%/5mm criterion; such result was improved to 99.9%/99.6% when 10%/10mm was adopted.

*Conclusion:*

The developed biologically guided deep learning method achieved post-20Gy $^{18}$FDG-PET image outcome predictions in good agreement with ground-truth results. With break-down biological modelling components, the outcome image predictions could be used in adaptive radiotherapy decision-making to optimize personalized plans for the best outcome in future.

rate of test images was 96.5%/92.8% using 5%/5mm criterion; such result was improved to 99.9%/99.6% when 10%/10mm was adopted.

*Conclusion:*

The developed biologically guided deep learning method achieved post-20Gy $^{18}$FDG-PET image outcome predictions in good agreement with ground-truth results. With break-down biological modelling components, the outcome image predictions could be used in adaptive radiotherapy decision-making to optimize personalized plans for the best outcome in future.

# 1. Introduction

Radiotherapy is a central component of the standard of care for many cancers. In the current era of Image-Guided Radiotherapy (IGRT), medical imaging plays a critical role in radiotherapy practice regarding patient assessment, treatment volume definition, on-board patient positioning, and outcome assessment[1]. In particular, imaging-based radiotherapy outcome assessment can capture early therapeutic responses for adaptive therapy to enhance radiotherapy efficacy[2]. In addition, long-term therapeutic outcomes from image-based analysis provide useful information in treatment intervention of each individual patient towards optimized cancer care[3]. Thus, medical imaging analysis for radiotherapy outcome assessment has become an irreplaceable component in precision medicine.

Technologies of medical imaging analysis have revolutionized image-based radiotherapy outcome reporting. Radiographic assessment of post-radiotherapy tumor morphological changes (i.e., Response Evaluation Criteria in Solid Tumors [RECIST]) was standardized to describe response to therapy[4]. Functional imaging modalities have now shifted outcome analysis from morphological description to physiological characterization. Positron emission tomography (PET) tracks the *in vivo* radioactive tracer distribution, for example estimating glucose metabolism ($^{18}$F-FDG) or measuring tissue hypoxia ($^{18}$FMISO)[5]. MR functional imaging, including dynamic contrast-enhanced MRI (DCE-MRI), diffusion weighted MRI (DWI), and diffusion tensor MRI (DTI), can measure tissue properties such as blood volume/perfusion[6], cellular density[7], and cell movement direction[8]. To non-invasively quantify *in vivo* physiology, functional imaging relies on mathematical models to extract quantitative parameters from phenotype image data. These mathematical models, which are often referred as mechanism-based models, describe complex physiological processes using basic biological theories and fundamental laws in physical/chemical interactions[9,10]. The derived parameters of mechanism-based models can serve as surrogates of individual physiology functions to facilitate developing a personalized therapeutic approach.

Treatment response assessment using functional imaging is often reported as post-treatment changes relative to pre-treatment baseline values. Image-based treatment outcome prediction, i.e., forecasting post-treatment image volumes before treatment initiation, has become an emerging topic in clinical oncology[11]. Potential clinical application of image-based treatment outcome prediction in radiotherapy is conceptually promising: given an individual's pre-radiotherapy image , post-radiotherapy image predictions could be available at the treatment planning stage. Guided by these predictions, clinicians could simulate alternative treatment plans, such as target delineation revision and plan parameter tuning (beam angle, energy selection, etc) for normal tissue sparing, and select a plan that predicts improved response to radiotherapy. This scenario can be applied to adaptive radiotherapy: the predicted intra-treatment images can be used to determine whether a revised radiotherapy plan would be advantageous. Additionally, when new intra-treatment image data are collected, the updated predictions can guide the adaptive planning strategy for optimal radiotherapy outcomes of individual patients[10]. Driven by the rapid growth of computation power, deep learning techniques have recently become a practical approach for image-based treatment outcome prediction[12-14]. However, few investigators have reported functional image outcome prediction in radiotherapy application. Aside from the colossal computational workload due to image dimension requirement, the current mechanism-based models focus on spatial decoding of physiology within an image volume; for outcome prediction, a mechanical-based model must incorporate patient-specific treatment information to simulate spatiotemporal physiology evolution during a treatment course. Although pilot studies have reported the feasibility of post-radiotherapy functional image outcome prediction using treatment information[15], the adopted deep learning network ignored the biophysical modelling and generated its prediction as a 'black box'; thus, the achieved prediction was reported at a fixed time point without any biological interpretation. Radiotherapy outcome prediction with biological modelling for time-series image prediction is an unmet need.

In this work, we design a biologically guided deep learning framework for intra-treatment $^{18}$FDG-PET image outcome prediction in response to oropharyngeal cancer intensity-modulated radiotherapy (IMRT). Based on the classic reaction-diffusion mechanism in disease progression, we propose a novel partial differential equation (PDE) as a biological model that incorporates spatial radiation dose distribution as a patient-specific treatment information variable. An encoder-decoder based convolutional neural network (CNN) is designed and trained to learn the proposed model, which governs the dynamics of tissue response to radiotherapy. Thus, the biological model learned by the CNN can generate a prediction of post-radiotherapy $^{18}$FDG-PET image outcome that reveals a time-series evolution.

## 2. Materials and Methods

*Biological Modelling*

We hypothesize that the SUV (Standardized Uptake Value) change in $^{18}$FDG-PET in response to radiation can be described in a reaction-diffusion system, which represents a family of mathematical models widely used in describing pattern formation and evolving densities in physical, ecological, and biological systems[16]. In the context of modeling tumor growth and therapeutic response dynamics, reaction-diffusion models have been applied to both pre-clinical and clinical works[9,17,18]. Disease progression in general can be summarized by Eq. (1) that describes the malignancy proliferation (reaction) and spread (diffusion)[10]:

$$U_t = \alpha \Delta U + \beta U \qquad (1)$$

where $U$ is the spatial distribution of disease (i.e., SUV intensity distribution in this work), and $U_t = \frac{\partial U}{\partial t}$ is the time derivative term describing the change of $U$ in time. The term $\alpha \Delta U = \alpha(\frac{\partial^2 U}{\partial x^2} + \frac{\partial^2 U}{\partial y^2})$ describes the spreading of abnormal cell activities, where $\alpha > 0$ is the diffusion coefficient. The linear term $\beta U$ represents the proliferation of localized malignancy. To incorporate tissue response to radiotherapy in the model in Eq. (2) we propose a new response term for the dose-induced changes of $U$,

$$U_t = \alpha \Delta U + \beta U + \mathcal{F}(DU) \qquad (2)$$

where $\mathcal{F}(DU)$ is an unknown operator that depicts $U$'s local response to radiotherapy. Here we assume that the response term depends on the product of $U$ and the radiotherapy plan's spatial dose distribution $D$. We also assume that the operator $\mathcal{F}$ depends on the local spatial information of $DU$ and we will use a convolutional neural network to learn this operator. Thus, Eq. (2) is the core time-dependent partial differential equation that models the post-radiotherapy biological response of abnormal tissue metabolism as SUV intensity (i.e., $U$) evolves in time.

*Deep Learning Design*

Formally, our problem is defined as follows: Given a set of pre- and post-radiation $^{18}$FDG-PET image pairs $\{(U_k^{pre}, U_k^{post})\}_{k=1,2,...m}$ and the imposed radiation dose distribution images $\{D_k\}_{k=1,2,...m}$, where $m$ is the total number of image pairs, our goal is to learn the unknown response operator $\mathcal{F}$ and coefficients α, β in the model Eq. (2) with the collected data of the form $\{(U_k^{pre}, U_k^{post}, D_k)\}_{k=1,2,...m}$. Accordingly, the learned model can predict a post-radiation $^{18}$FDG-PET image $U_k^{post}$ given pre-radiation image $U_k^{pre}$ and the associated spatial dose distribution $D_k$. In addition, since the learned model describes the evolution dynamics of $U_k$ between the two states $U_k^{pre}$ and $U_k^{post}$, multiple frames of $U_k$ between $U_k^{pre}$ and $U_k^{post}$ can be simulated to study the intermediate stages of disease progression.

While a large body of work has focused on solving reaction-diffusion models like Eq. (2), i.e., finding $U$ based on known coefficients and operators, little research has been devoted to the inverse problem of learning the model's coefficients and operators from observed $U$ data. The numerical treatments of the inverse problem are typically complicated, as the observed data usually cannot provide sufficient information to determine a unique model, and regularizations are needed to produce meaningful model estimates. As such, we propose a deep neural network framework to learn the model in Eq. (2) from $^{18}$FDG-PET images taken before and after radiation. Applying the forward-Euler method on the PDE in Eq. (2), we obtain the discretized update rule:

$$U^{n+1} = U^n + h\alpha\Delta U^n + h\beta U^n - h\mathcal{F}(DU^n) \qquad (3)$$

where $h$ is the time step, $U^n$ is the approximate solution of the state $U$ at time $t_n = nh$, and the Laplacian operator $\Delta$ can be approximated by a discrete operator $D_{xy}^2$ represented by a nine-point stencil[19]:

$$D_{xy}^2 = \begin{pmatrix} 1/4 & 1/2 & 1/4 \\ 1/2 & -3 & 1/2 \\ 1/4 & 1/2 & 1/4 \end{pmatrix} \qquad (4)$$

A deep neural network $\mathcal{N}_{\mathcal{F}}$ is designed to approximate the response operator $\mathcal{F}$:

$$\mathcal{N}_{\mathcal{F}}: \psi \rightarrow \mathcal{N}_{\mathcal{F}}(\psi; \theta) \quad (5)$$

where θ represents the free parameters. For simplicity, we assume that the operator $\mathcal{F}$ only depends on $\psi = DU$, the product of the dose distribution $D$ and the $^{18}$FDG-PET image state variable $U$. A diffusion-proliferation operator $\mathcal{G}$ is used to combine both the diffusion and proliferation terms with undetermined coefficients $\alpha$ and $\beta$:

$$\mathcal{G}(U^n) = \alpha D_{xy}^2 U^n + \beta U^n \quad (6)$$

Given a group of three images consisting of the initial state $^{18}$FDG-PET image $U_k^0 = U_k^{pre}$ at $t = 0$ prior to radiation, dose distribution map $D_k$, and the ground-truth final state $^{18}$FDG-PET image $U_k^{post}$ at $t = T$ (post-radiation), from Eqs. (3)-(6) we obtain the intermediate states of $U_k^{n+1}$ by:

$$U_k^{n+1} = U_k^n + h\mathcal{G}(U_k^n) + h\,\mathcal{N}_{\mathcal{F}}(D_k \circ U_k^n; \theta) \quad (7)$$

for $n = 0, 1, \ldots N_t - 1$. Here, $D_k \circ U_k^n$ represents the element-wise product of the dose distribution map $D_k$ and the $^{18}$FDG-PET image $U_k^n$ at the time step $t_n$, $N_t$ is the total number of steps and the step size $h = T/N_t$. As a feasibility study, we consider the final time $T = 1$ in this work.

The similarity between the predicted post-radiation $^{18}$FDG-PET image $U_k^{N_t}$ and the associated ground-truth image $U_k^{post}$ is defined based on the $l_2$ norm loss function:

$$\mathcal{L}(\theta) = \frac{1}{m} \sum_{k=1}^{m} \left\| U_k^{N_t} - U_k^{post} \right\|_2^2 \quad (8)$$

where $m$ is the number of samples. By minimizing $\mathcal{L}(\theta)$, the deep neural network can learn the weights $\theta$ that characterize the response operator $\mathcal{F}$ and the undetermined coefficients $\alpha$ and $\beta$.

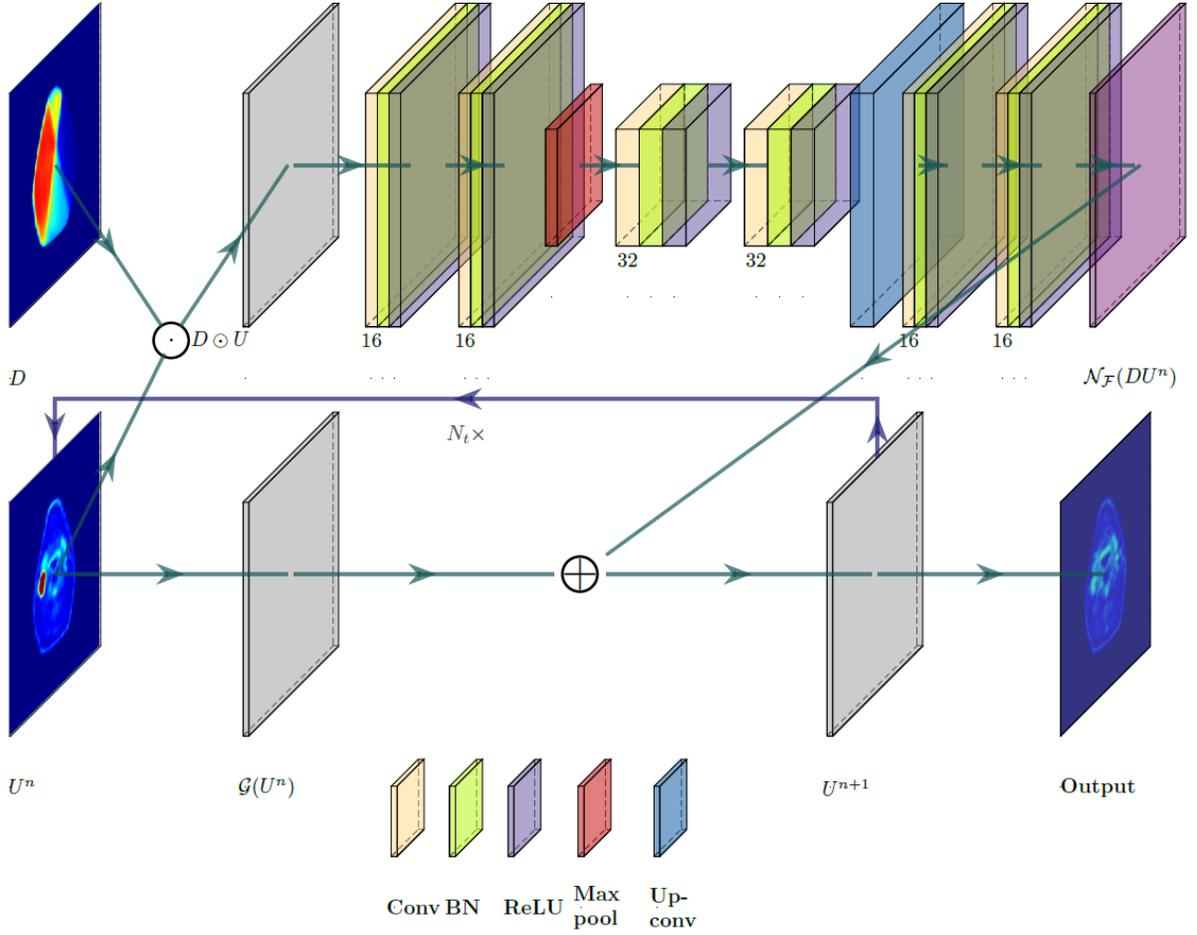

Figure 1: A PDE-informed deep neural network design. Layers are color-coded by operations with associated feature numbers.

Figure 1 illustrates the designed deep neural network architecture. The network's input space is composed of pre-radiation [18]FDG-PET image $U^{pre}$ and planned dose distribution map $D$ as a set. The network is split into two branches: one that uses a convolutional neural network to learn the response operator $\mathcal{N}_{\mathcal{F}}(DU^n)$, and the other one with only two trainable parameters to apply the diffusion-proliferation operator $\mathcal{G}$ (in Eq. (6)) on $U^n$. Specifically, the second branch of the network architecture mimics the traditional finite difference method and applies the discrete Laplacian operator and the linear operator on $U^n$ with predicted $\alpha$ and $\beta$. Both branches are then merged by the rule in Eq. (7), which feeds the output $U^{n+1}$ forward to the next cycle. This process is then repeated for $N_t$ time steps to generate a predicted post-radiation [18]FDG-PET image, which will be compared against the ground-truth post-radiation [18]FDG-PET image.

The branch that learns the response operator $\mathcal{N}_\mathcal{F}(DU^n)$ consists of a total of 7 convolutional layers and is built upon U-Net's encoder-decoder architecture[20]. The architecture consists of a contracting path that extracts sufficient semantic context from $D \circ U^n$ and a symmetric expanding path that produces the up-sampled output. The contracting path starts with two applications of 3×3 convolutions, each followed by a batch normalization layer and a ReLU operation. Then a 2×2 max pooling operation is performed for down-sampling where the number of feature channels is doubled. Then another two 3×3 convolutions operations are applied, each followed by a batch normalization and a ReLU activation. The expanding path consists of an up-sampling of the feature map, followed by two 3×3 convolutional layers, again with batch normalization and ReLU operations. Finally, a 1×1 convolution is applied to map the 16-component feature to a single feature channel that reconstructs the transformed image corresponding to $\mathcal{N}_\mathcal{F}(DU^n)$.

*Patient Data and Network Training*

In this work, 64 eligible oropharyngeal cancer patients who received curative-intent IMRT in our department were retrospectively studied under an IRB-approved $^{18}$FDG-PET imaging study. Prior to treatment initiation, each patient underwent a $^{18}$FDG-PET/CT scan for target delineation. After 20Gy delivery of a planned 70 Gy at 2 Gy/daily fraction, each patient underwent a second $^{18}$FDG-PET/CT scan as an intra-treatment evaluation for consideration for adaptive planning. These post-20Gy $^{18}$FDG-PET acquisitions were treated as the post-radiation scans in the modelling.

All $^{18}$FDG-PET/CT exams were acquired by a PET/CT scanner (Siemens, Erlangen, Germany) in our department. PET acquisitions were performed using 400 × 400 matrix size in a standard field of view (FOV) of 54 cm, and slice thickness was 2 mm. CT acquisitions were performed using 512 × 512 matrix size in an extended FOV of 65 cm, and slice thickness was 3 mm. PET images were reconstructed by the ordered subset expectation maximization (OSEM) algorithm with attenuation corrections using the CT acquisition information. The post-20Gy $^{18}$FDG-PET/CT images were registered to the corresponding pre-radiation images using Velocity$^{TM}$ software (Varian Medical Systems, Palo Alto, CA, United States). Registrations started with rigid bony structure alignment, and multi-pass deformable registration algorithm was adopted to improve soft tissue alignment near the anterior body surface. In the process of IMRT planning, all treatment plans were optimized and calculated using Eclipse™ treatment planning system (Varian Medical Systems, Palo Alto, CA, United States) with a 2.5mm dose calculation grid size. All $^{18}$FDG-PET images and spatial dose distribution maps of 20Gy treatment were resampled to the CT simulation image grid size.

Of all 2D $^{18}$FDG-PET axial images, those with high $^{18}$FDG uptake ($SUV_{max}$ > 1.5) in the pre-radiation acquisition were selected. Overall, 718 axial slices collected from 38 patients were used for neural network training, 233 axial slices from 13 patients were used for validation, and 230 axial slices from 13 patients were used for independent tests. During the neural network training, the loss function was defined as based on the $l_2$ norm in Eq. (8). Gradient updates were computed using batch sizes of 10 samples, and batch normalization was performed after each convolutional layer. The training utilized the Adam optimizer for up to 400 epochs, while an early stopping strategy on the loss function evaluated on the validation samples was adopted with a patience of 100 epochs. The overall training time was about 15 minutes in TensorFlow environment using a NVIDIA TITAN™ Xp graphic card.

*Evaluation*

The accuracy of post-20Gy $^{18}$FDG-PET image prediction was evaluated using 230 axial slices' results from 13 test patients. The prediction results were first visually inspected with morphological features as qualitative evaluation. SUV mean values in high-uptake regions determined by Otsu's method[21] were quantitatively compared with the ground-truth results. Pixel-to-pixel SUV numerical differences was evaluated by Gamma tests within the body region[22]. Multiple Gamma tests with different SUV difference tolerances and distance-to-agreement (DTA) tolerances were performed. While Gamma Index < 1 was considered as acceptable pixel-wise results, Gamma Index passing rates, i.e., the percentage of pixels with Gamma Index < 1, were reported as summarizing metrics.

## 3. Results

Figure 2 shows an example case of post-20Gy $^{18}$FDG-PET image outcome prediction. As seen in the pre-radiation $^{18}$FDG-PET image, SUV hotspots with clear edge were found on the patient's right side. After 20Gy delivery shown by the bilateral side dose distribution in *D*, the ground-truth post-20Gy $^{18}$FDG-PET image results demonstrated good therapy response with reduced hotspot sizes and decreased SUV intensities. The predicted $^{18}$FDG-PET image captured the overall morphological appearance in the ground-truth results without noticeable artifact marks. Two hotspots corresponding with nodal disease were found in the prediction image at the same locations. The hotspots' sizes and SUV intensities were comparable, though the anterior hotspot intensity appeared to be lower than the ground-truth result. In the breakdown illustration of biological model terms in Eq. (3), the diffusion term demonstrated overall uniform intensity distribution around 0 except hotspot regions; the core regions in hotspots had negative diffusion intensities, which suggested a spatial retraction of abnormal metabolism. The proliferation term had a similar appearance to the pre-radiation $^{18}$FDG-PET image. The dose response term indicated an elevated intensity region that corresponds to the anterior SUV hotspot; this suggests that the anterior SUV hotspot had a better response to 20Gy than the posterior SUV hotspot, which had limited intensity in the dose response map. The other areas within the body had close-to-zero dose response intensity, while low intensities were found near body surface. The Gamma Index map showed good quantitative pixel-to-pixel SUV comparison between ground-truth and predicted post-20Gy $^{18}$FDG-PET images using 5%/10mm criterion.

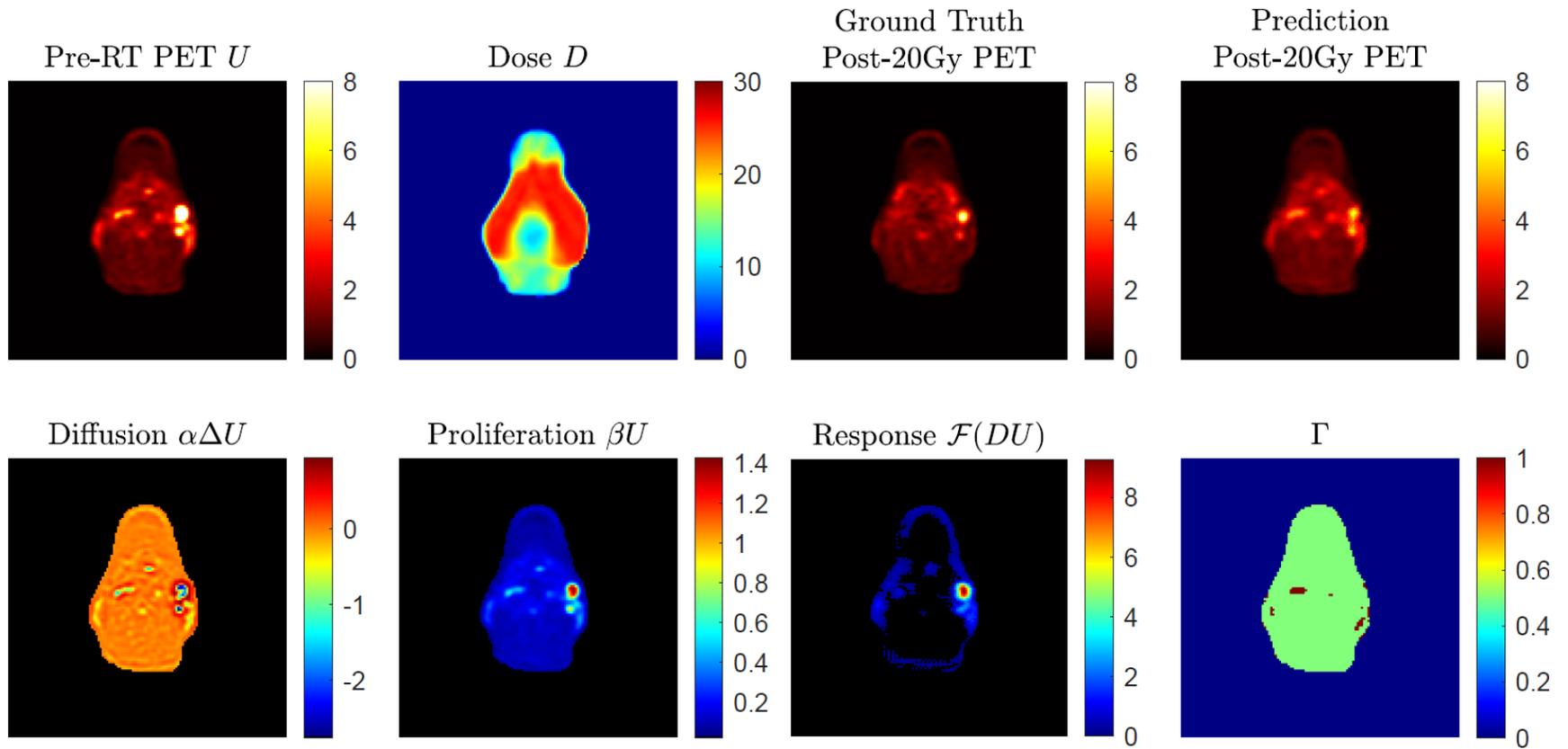

Figure 2: An example of post-radiotherapy $^{18}$FDG-PET image outcome with given pre-radiation $^{18}$FDG-PET image and dose distribution map D, with a breakdown of predicted biological effects (diffusion, proliferation, and response) in Eq. (2). The 2D Gamma ($\Gamma$) test result was obtained through acceptance criteria 5%/10mm.

Figure 3 shows an example of intermediate stage $^{18}$FDG-PET image outcome predictions as biological model solutions from pre-radiation result at $t = 0$ to post-20Gy prediction at $t = 1$. In general, the four predicted $^{18}$FDG-PET images demonstrated a reasonable image state transition from $t = 0$ to $t = 1$ without abrupt changes. While the majority of normal tissue maintained steady SUV intensities during the presented time evolution, the SUV hotspot corresponding to the primary oropharyngeal tumor had shrinkage at its posterior boundary with slightly reduced intensity. Compared to the ground-truth post-20Gy $^{18}$FDG-PET image, the prediction image at $t = 1$ captured the SUV hotspot's morphological features, particularly at its posterior boundary.

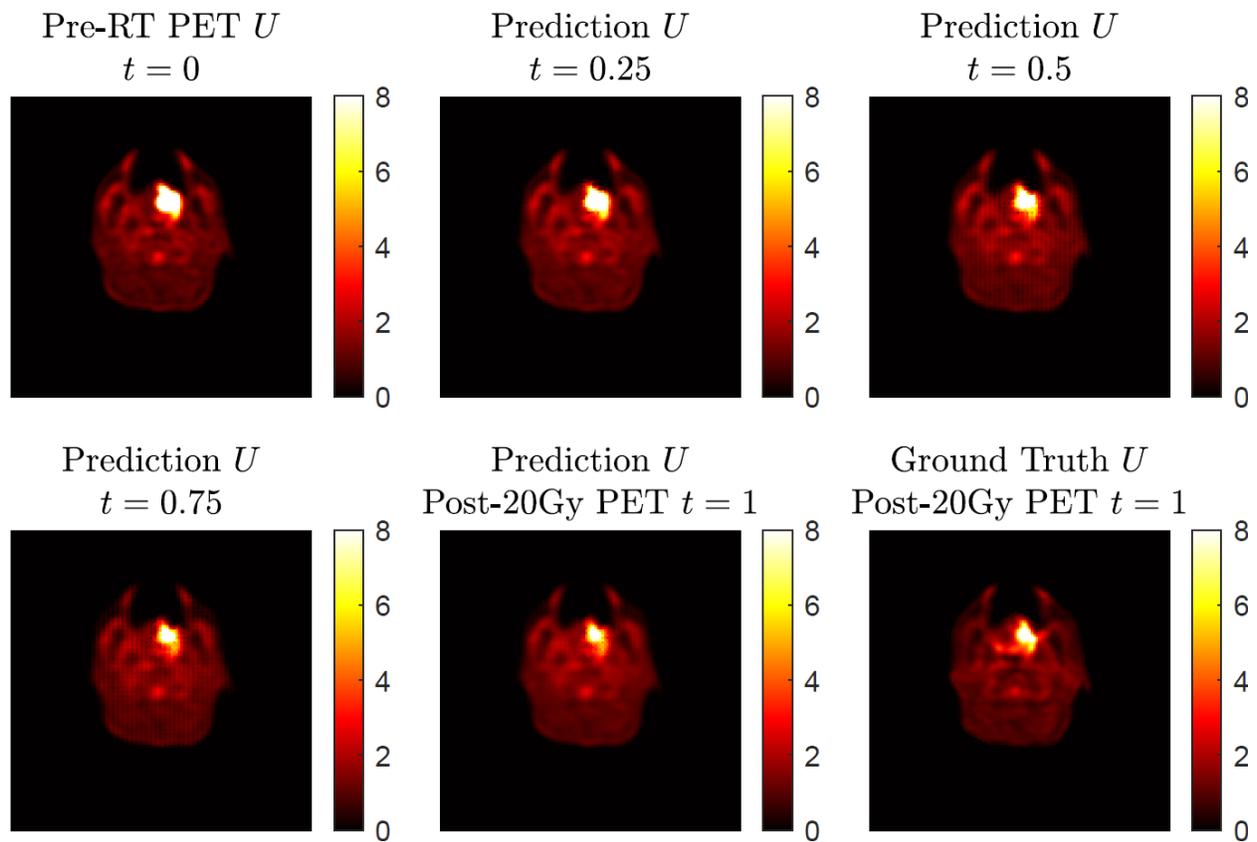

*Figure 3: An example of $^{18}$FDG-PET image outcome prediction based on a four-step execution with step size h= 0.25 showing the predicted transition from pre-radiation U at t= 0 to the predicted post-radiotherapy U at t = 1. The ground-truth post-20Gy $^{18}$FDG-PET image is included for comparison at t =1.*

In the test patient cohort, the SUV mean value of high-uptake regions in post-20Gy predicted images were 2.45±0.25, which was slightly lower than ground-truth results (2.51±0.33, $p = 0.015$). Gamma Index passing rate results of all testing axial slices are summarized in Figure 4. When 5%/5mm Gamma criterion was adopted, the median 2D Gamma passing rate was 96.5%. Using looser Gamma criteria, the passing rate results improved (5%/10mm: median 99.2%, average 97.6%; 10%/5mm: median 99.5%, average 98.6%). The highest median passing rate was 99.9% (average = 99.6%) when 10%/10mm was used.

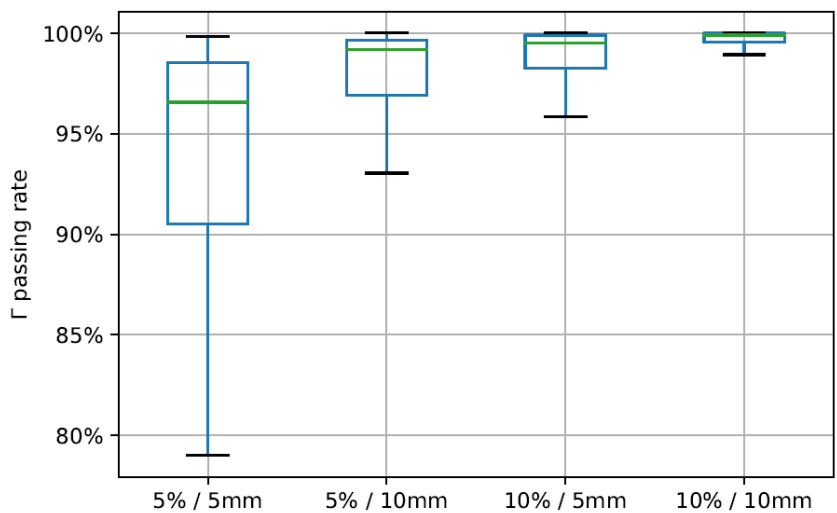

*Figure 4: Gamma Index passing rate summary with different gamma criteria. Green line positions represent median value and box represent 25%/75% percentile with whiskers indicating 5%/95% percentile.*

## 4. Discussion

In this work, we successfully demonstrated the design of a biological model guided deep learning framework for post-20Gy $^{18}$FDG-PET image outcome prediction in a unique cohort of patients undergoing IMRT for oropharyngeal cancer. One of the key innovations in this work is the biological model in Eq. (2), which was hypothesized as the mathematical equation that governs the post-radiation SUV change. The model was derived from the classic reaction-diffusion system, which has been utilized in many works of tumor growth and disease progression modelling[23-25]. Although applying reaction-diffusion models to $^{18}$FDG-PET image analysis (particularly to head and neck cancer) is less reported, some exploratory studies have demonstrated the validation of reaction-diffusion-type models in intra-cranial PET image modelling[26]. Compared to the original reaction-diffusion models, the newly introduced dose response term in Eq. (2) was hypothesized as a semantic component of dose-induced SUV image state changes. Adding additional terms in reaction-diffusion family models to account for therapeutic effect has been reported before in breast, lung, and pancreatic cancer studies[27-29]; nevertheless, our approach of using spatial dose distribution in biological modelling is a novel design. Compared to the use of prescription dose levels for outcome assessment/prediction in many studies, the adoption of spatial dose distribution maintained heterogeneous radiation deposition information at pixel level, which may be a more accurate approach for image-based outcome prediction.

As a deep learning approach, a CNN was designed to learn the proposed biological model. Partial differential equations with known coefficients and operators can be solved by various numerical methods such as finite difference methods, finite element methods, and spectral methods. In the scientific computing field, solving partial differential equations using CNN in complex systems has become popular for efficiency and accuracy[30]. In this work, the use of CNN is necessary to learn the dose response term $\mathcal{F}$ in Eq. (2), which is an unknown operator that is assumed to be related to the product of spatial dose distribution and $^{18}$FDG-PET image variable ($DU$); without an analytical expression, it is difficult to approximate the operator $\mathcal{F}$ by classic numerical treatments of inverse problems. The proposed CNN in Figure 1 revealed the dose response term $\mathcal{F}(DU)$ as a whole, while the detailed mechanism of $DU$'s contribution of $^{18}$FDG-PET image prediction remains unclear. Inspired by the classic encoder-decoder U-net implementation, the CNN architecture in Figure 1 was dedicated for the problem in Eqs. (3)-(7); with the loss function defined in Eq. (8), the training process had a fast convergence (Figure 5). It would be of interest to further study the operator $\mathcal{F}$ for its analytical expression and possible biological explanations. Such works require more advanced mathematical theories supported by experimental data, preferably as *in vitro* implementations, to validate analytical designs as a biological model calibration process[10].

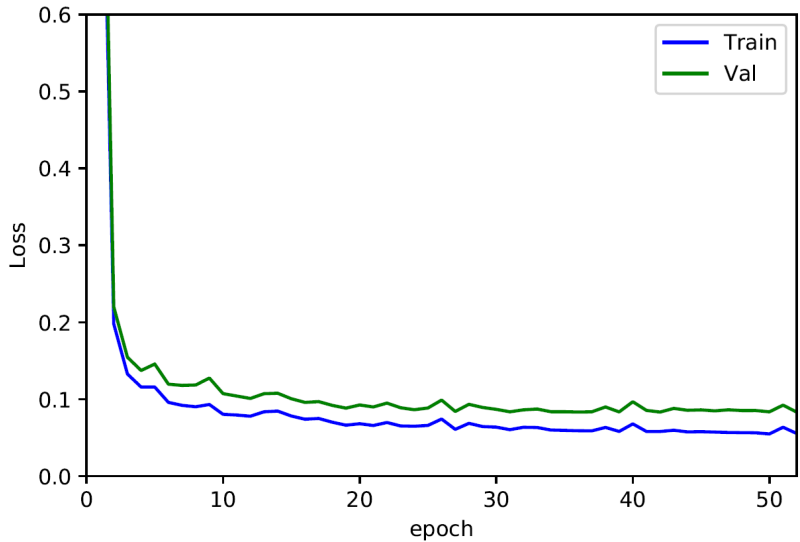

*Figure 5: Training (Blue) and validation (green) loss history*

Based on the Gamma test results in Figure 4, the achieved $^{18}$FDG-PET image predictions showed good agreement with ground-truth images. As a common quality assurance method in radiotherapy practice, Gamma analysis accounts for both intensity differences and systematic shifts in image prediction error. The Gamma test criteria need to consider multiple uncertainty sources in data processing and clinical preferences. For instance, the dose response term results in Figure 2 indicated very small but non-zero intensity values near body surface, especially at anterior skin regions. While other normal tissues demonstrated very limited dose response, the observed skin regions' dose response may be noisy results related to deformable image registration uncertainties, which was mainly determined by patient weight loss during the radiotherapy course[31]. Radiotherapy margin formulism that models treatment margin statistics should also be weighted in image prediction evaluation[32]. In addition to these two potential factors, the adopted Gamma test criteria has incorporated many other factors, including SUV's intrinsic uncertainty, PET image acquisition resolution, PET-CT QA protocol, and SUV-based metabolic volume definition.

The current biological model was implemented in a 2D fashion on axial images. For each test patient, the post-radiation $^{18}$FDG-PET image predictions were generated slice-by-slice to approximate volumetric rendering. In theory, the biological model in Eq. (2) and the demonstrated deep neural network could be implemented as 3D in spatial domain; however, the computation workload for 3D implementation would be much larger. A more prominent problem of 3D modelling is the limited data sample size. Currently, 1181 high $^{18}$FDG uptake axial slices were collected from 64 patients, and these slices were assigned to neural network training/validation/tests following 60%/20%/20% ratio. While axial slices can be treated as independent data samples, 3D-based modelling can only utilize 64 volumes as independent data samples from 64 patients; this would be a very limited data sample size for deep learning applications. Indeed, a large patient cohort size is optimal for deep learning applications; however, like most other similar works, limited resources for intra-treatment imaging and the retrospective nature of this study make it challenging to find sufficient eligible cases for 3D-based image outcome prediction. Experiments using small animals are planned future development of deep learning in image outcome prediction. Further investigation of the biological interpretation of the learned dose response term may also lead to improved mathematical modeling for this problem.

As a feasibility study, the current results showed that the achieved post-20Gy $^{18}$FDG-PET image outcome prediction had good agreement with ground-truth results. Post-20Gy $^{18}$FDG-PET has been demonstrated as informing surrogates of recurrence-free survival and overall survival of human papillomavirus (HPV)-related oropharyngeal cancer[33]. In a potential clinical application scenario, the current framework would allow a physician to determine if an $^{18}$FDG-PET scan after

20Gy radiation would facilitate improved adaptive radiotherapy clinical decision making. One crucial step towards this clinical application scenario is to verify the models' responses to different radiation therapy strategies. The current patient cohort from a clinical study received a uniform treatment regimen; thus, the developed model may not capture certain individual reactions after a drastically different radiotherapy approach. For deep learning developments, it would be ethically challenging to collect patient data with intentional treatment variations. Following the small animal experiments discussed above, with dedicated imaging platforms and radiotherapy machines, one can generate post-radiation samples with heterogeneous treatment strategies in multiple imaging sessions. Such experiments may provide valuable opportunities for studying biological models for improved deep learning intelligibility.

## 5. Conclusion

In this work, we developed a biological model guided deep learning method for post-radiation $^{18}$FDG-PET image outcome prediction. The proposed biological model incorporates spatial radiation dose distribution as a patient-specific variable, and a novel CNN architecture was implemented to predict post-radiotherapy $^{18}$FDG-PET images from pre-radiation results. Current results demonstrate good agreements between post-20Gy predictions and ground-truth results in a cohort of patients with oropharyngeal cancer. Future developments of current methodology design in small animal experiments will enhance the applicability of image outcome prediction in clinical practice.

**Figure Legends**

Figure 1: A PDE-informed deep neural network design. Layers are color-coded by operations with associated feature numbers.

Figure 2: An example of post-radiotherapy $^{18FDG-PET}$ image outcome with given pre-radiation $^{18FDG-PET}$ image and dose distribution map D, with a breakdown of predicted biological effects (diffusion, proliferation, and response) in Eq. (2). The 2D Gamma (Γ) test result was obtained through acceptance criteria 5%/10mm.

Figure 3: An example of $^{18FDG-PET}$ image outcome prediction based on a four-step execution with step size h= 0.25 showing the predicted transition from pre-radiation U at t= 0 to the predicted post-radiotherapy U at t = 1. The ground-truth post-20Gy $^{18FDG-PET}$ image is included for comparison at t =1.

Figure 4: Gamma Index passing rate summary with different gamma criteria. Green line positions represent median value and box represent 25%/75% percentile with whiskers indicating 5%/95% percentile.

Figure 5: Training (Blue) and validation (green) loss history

# Acknowledgments

H. Ji was partially supported by the Simons Foundation Math+X investigator award number 510776 and the National Science Foundation under grant NSF DMS-1952339.